\documentclass{ws-ijmpd}

\usepackage[super,compress]{cite}
\usepackage{graphicx,color}
\def\cblue{\color{blue}}

\usepackage{hyperref}
\hypersetup{colorlinks,urlcolor=black,citecolor=black,linkcolor=black,filecolor=black}
\usepackage{breakurl}

\begin{document}

\markboth{Gustavo O. Heymans, Nami F. Svaiter and Gastão Krein}
{Analog Model for Euclidean Wormholes Effects}

%
%

\title{Analog Model for Euclidean Wormholes Effects}  

\author{Gustavo O. Heymans$^*$ and Nami F. Svaiter$^{\dagger}$}

\address{Centro Brasileiro de Pesquisas F\'{\i}sicas,
Rua Xavier Sigaud 150, Rio de Janeiro, 22290-180 Rio de Janeiro, RJ, Brazil.\\
$^*$olegario@cbpf.br\\
$^{\dagger}$nfuxsvai@cbpf.br}

\author{Gastão Krein}

\address{Instituto de F\'{i}sica Te\'orica, Universidade Estadual Paulista, 
Rua Dr. Bento Teobaldo Ferraz, 271, Bloco II, 01140-070, S\~ao Paulo, SP, Brazil \\ 
gastao.krein@unesp.br}

\maketitle
\begin{abstract}
Using results of statistical field theory 
for systems with an anisotropic disorder, we present an analog 
model for Euclidean wormholes and topological fluctuation effects in a Riemannian space $\mathcal{M}^\mathrm{d}$. 
The contribution of wormholes and topological fluctuations to the Euclidean gravitational functional 
integral is modeled by quenched randomness defined in the $\mathbb{R}^{\mathrm{d}}$ 
manifold. We obtain a disorder-averaged free energy by taking the average over all the realizations 
of the random fields. In the scenario of topology fluctuation, there appears a superposition of infinite 
branes that contribute to the physical quantities. All topology fluctuations can be understood as two distinct 
kinds of Euclidean wormholes: wormholes confined to one brane, and wormholes connecting different branes. 

\vspace{1.5cm}
\noindent{*Honorable Mention in Gravity Research Foundation 2023 Awards for Essays on Gravitation}
\end{abstract}


\ccode{PACS numbers:}


\section{Introduction}	

Relativistic quantum field theory 
is defined in the four-dimensional Minkowski space with a causal and metric structure associated with
the Poincaré group generated by the Lorentz transformations and translations. In the operatorial 
approach of the theory, quantum fields $\phi$ are operator-valued distributions~$\phi(f)$ smeared 
over smooth test functions~$f$ with compact support acting as unbounded operators in a Hilbert 
space. The principles underlying a quantum field theory are the probabilistic interpretation of 
expectation values, special relativity, and locality. Locality means that two smeared $\phi(f)$ 
and $\phi(g)$ field operators commute for smearing functions $f$ and $g$ with spacelike separated 
supports. 
 
The program of describing the gravitational field using quantum theory faces many conceptual 
difficulties, mainly related to causality and locality. Quantum field theory formulated on a 
classical gravitational background spacetime is an intermediate step toward such a 
program~\cite{davies,lectford,Hollands:2014eia}. A problem that permeates this approach is the absence of 
a specific vacuum state associated with matter fields in a generic spacetime; but in  
globally hyperbolic spacetimes, it is circumvented by the use of Hadamard states. To go further, one 
can discuss the effects of the fluctuations of the metric fields over the quantum matter fields. 
One can show that a bath of gravitons in a squeezed state induces fluctuations of light 
cones~\cite{Ford:1994cr,Ford:1996qc,Ford:1997zb}. More than ten years ago G. 
Menezes and two of us proposed an analog model for fluctuating light cones induced by quantum gravity 
effects~\cite{Krein:2010ee}. The model builds on the fact that acoustic waves in a disordered medium 
propagate with a random speed of sound. Further studies discussing analog models can be found 
 in Refs.~[\refcite{Arias:2011yg,Bessa:2012fs,Ford:2012xk,Arias:2013mza}]. In the present 
, we build on similar ideas to propose an analog model for Euclidean wormhole 
effects on a real scalar field. 

In recent years, there has been a growing perception~\cite{Giddings:2022jda} that 
long-distance physics issues are as important in quantum gravity as the most-discussed 
short-distance physics issues. A central, open question in this respect is: How does nonlocality 
of quantum gravity affect the expectation value of a measurable observable? 
Going back to the local quantum field theory, from the expectation values 
of operator products in Minkowski spacetime one can define the Schwinger functions, 
which are the corresponding analytically continued vacuum expectation values
in Euclidean space~\cite{Wick:1954eu,Schwinger:1958qau,nakano,e3}. This possibility of a Euclidean
field theory is supported by the Osterwalder-Schrader reconstruction theorems~\cite{Osterwalder:1973dx,Osterwalder:1974tc}. 
In the functional integral formalism~\cite{Gelfand:1959nq} in Euclidean space, the Schwinger functions are moments of a measure 
in the functional space of classical fields. In such 
a functional integral scenario, one can discuss topology fluctuations and wormholes~\cite{Hawking:1988ae,
Coleman:1988tj,Klebanov:1988eh,Preskill:1988na}, as quantum gravity features particularly relevant to the 
issue of quantum coherence loss in Hawking black hole evaporation.

The basic feature of wormholes in the Euclidean 
field theory is the existence of nonlocal physics in a connected manifold or geometry that connects disconnected boundaries. The contribution to the 
free energy from these connected topologies was discussed in Ref.~[\refcite{Engelhardt:2020qpv}]
using the replica trick. The replica trick provides a convenient way to compute 
averages of the free energy (the log of the partition function)~\cite{re}. 
 In a related study, Ref.~[\refcite{okuyama}] proposes an integral representation of $\ln{x}$  to 
compute the free energy of spacetime $D$-branes. The author of that reference argues that the bulk gravity picture of such an integral representation involves 
wormholes connecting multiple asymptotic boundaries. Replica wormholes also play a role in the computation of von Neuman entropy of Hawking radiation~\cite{Dong:2016fnf}. 

Two of the most fundamental questions facing Euclidean quantum gravity are the 
following: 1) What is the empirical support of this mathematical formalism of Euclidean quantum gravity? 
2) What are the physical effects of topological fluctuations on 
the Euclidean quantum fields? In the absence of cosmological experiments, we propose a condensed matter analog 
model that might bring insight about those questions. We propose an analog model for 
topological fluctuations on Euclidean fields based on external disordered fields  described
by a statistical field theory. In a statistical field theory, 
the  dynamical degrees of freedom are random variables distributed according to the Gibbs measure. 
In the context of such a field theory, a large class of quenched-disordered systems can be 
modeled by coupling the system's field degrees of freedom to 
an external field with a prescribed probability distribution. 
Averaging the free energy
by integrating over the disorder fields defined by some probability distribution results
in an effective action for the matter fields. 
It is a well-known fact~\cite{kirkpatrick,belitz1,Heymans:2022sdr,sachdevbook} 
that at low temperatures, or for an anisotropic disorder, 
disorder averaging induces nonlocal terms in the resulting
 effective action.
On the other hand, for systems at high temperatures, in which thermal fluctuations dominate, the corresponding effective action contains only local terms. 
A convenient way to deal with the quenched disorder averaging
is the distributional zeta function method~\cite{distributional,distributional2}, a method
based on a functional series representation of the quenched free energy. From  such a functional series, 
one is able to find the mean energy and mean entropy of disordered systems with induced nonlocal contributions in the effective action. 
 A scenario of matter fields in the presence  of topology fluctuations and wormholes 
(nonlocal action) then leads to the interpretation that the functional series represents a 
superposition of many branes with different configurations of Euclidean wormholes connecting them.

\section{Wormholes in a Riemannian Manifold}

We briefly discuss matter fields in a generic Riemannian manifold. Suppose a compact manifold 
of Riemannian signature $\mathcal{M}$. The space of fields are the space 
$C^{\infty}(\mathcal{M}, \mathbb{R})$ of smooth functions defined in $\mathcal{M}$. 
Let $S: C^{\infty}(\mathcal{M}, \mathbb{R}) \to \mathbb{R}$ be an action functional of the 
gravitational~$g$  and matter~$\phi$ fields. Using a functional measure for the 
gravitational and matter fields, the partition function is given by:
\begin{equation}
    Z = \int [dg][d\phi]\, e^{ - S(g) - S(\phi)}  
\end {equation}
where $S(g)$ and $S(\phi)$ are respectively gravitational field and matter field actions.
For simplicity, we take a single scalar field to represent the matter degrees of freedom. 
The gravitational field action is given by:
\begin{equation}
    S(g) = - \frac{1}{16\pi G} \int _{\mathcal{M}} d^{\mathrm{d}}x \, \sqrt{g} \, (R - 2\Lambda) 
    - \frac{1}{8\pi G} \int_{\partial \mathcal{M}} \, K d^{\mathrm{d}-1} \Sigma + C.
    \label{grav-action}
\end{equation} 
\noindent
As usual, $g= \mathrm{det}(g_{ij})$, $G$  is Newton's constant, 
$R$  the Ricci-scalar, $\Lambda$ the cosmological constant, $K$ 
the trace of the second fundamental form on the boundary, and $C$ a constant that can be tuned to achieve a convenient on-shell configuration, 
e.g. in flat space $S(g)=0$. For the matter field action we take:
\begin{align} 
    S(\phi) & = \frac{1}{2} \int  d^{\mathrm{d}}x \, \sqrt{g} \, \phi(x)\left(-\Delta + m^2\right)\phi(x)+ \frac{\lambda_0}{4} \int d^{\mathrm{d}}x \, \sqrt{g} \, \phi^4(x).
\end{align}

Many 
authors~\cite{Coleman:1988tj,Klebanov:1988eh,Preskill:1988na,
Giddings:1987cg} emphasized that the effects of wormholes and topology 
fluctuations are encoded in a nonlocal matter-field contribution to the 
Euclidean partition function, namely 
\begin{equation}
    Z = \int [dg] [d\phi]\exp\left[-S(\phi,g) 
    + \frac{1}{2}\int d^{\mathrm{d}}x\int d^{\mathrm{d}} y \sum_{i,j} \phi_i(x) \, C_{ij}(x,y) \, \phi_j(y)\right].
\label{klebanov}
\end{equation}
in which $C_{ij}(x,y)$ encodes the space nonlocality, in which each pair $i,j$ 
represents a wormhole{\cblue{.}} In the next section, we show that such a nonlocal 
term arises naturally in a matter system in the presence of disorder. 

As mentioned above, in a Euclidean quantum gravity 
scenario, many authors have been stressing the necessity of performing the average 
of the free energy, or generating functional of connected correlation 
functions of the system~\cite{Engelhardt:2020qpv,okuyama}. There are many ways to 
evaluate the quenched free energy. In this, we use the distributional 
zeta-function method~\cite{distributional, distributional2}.

\section{Field Theory in a Quenched Disordered Background}\label{sec:zeta}

Here we present a short review of quenched disorder coupling 
to a matter scalar field. We assume that the action functional in the presence 
of the disorder field is given generically by
\begin{equation}
S(\phi,h) = S(\phi) + \int d^{\mathrm{d}}x\,h(x)\phi(x),
\end{equation}
where  $S(\phi)$ is some action with fields $\phi(x)$ that for now we leave unspecified, 
and $h(x)$ is the disorder field. 
In a general situation, a disordered medium can be modeled by a real random field $h(x)$ in $\mathbb{R}^{d}$  
with $\mathbb{E}[h(x)]=0$ and nonvanishing covariance  $\mathbb{E}[h(x)h(y)]$, where $\mathbb{E}[...]$ 
specifies the mean over the ensemble of realizations of the disorder.

Let $Z(j,h)$ be the disorder generating 
functional of correlation functions, $i.e.$ the generating functional of 
correlation functions for one disorder realization:
\begin{equation}
Z(j,h)=\int[d\phi]\,\, \exp \left[-S(\phi,h) + \int d^{\mathrm{d}}x \,j(x)\phi(x)\right],
\label{eq:edisorderedgeneratingfunctional}
\end{equation}
with $j(x)$ being an external source field. As in the pure system case 
($i.e.$ $h=0$), one can define a disorder generating functional of connected 
correlation functions, $i.e.$, generating functional of connected correlation 
functions for one disorder realization, $W(j,h)=\ln Z(j,h)$, which is nothing else than 
the free energy of the system for one disorder realization. From one disorder 
realization, one can perform the quenched average free energy, performing the 
average over the ensemble of all realizations of the disorder. The average of the  free energy is then given by:
\begin{equation} 
\mathbb{E}\bigl[W(j,h)\bigr] = \int\,[dh]P(h)\ln Z(j,h), 
\end{equation}
where $[dh]=\prod_{x} dh(x)$ is a functional measure, and $P(h)$ a probability 
distribution of the disorder. For a generic $P(h)$, using the distributional zeta-function 
method one obtains
\begin{equation}
\mathbb{E}\bigl[W(j,h)\bigr] = \sum_{k=1}^{\infty} c_k \,\mathbb{E}\,[(Z(j,h))^{\,k}] 
- \ln(b) - \gamma+R(b,j),
\label{eq:logz}
\end{equation}
where $c_k = \frac{(-1)^{k+1}b^k}{k \,k!}$, $b$ is a
dimensionless arbitrary constant, $\gamma$ the Euler-Mascheroni 
constant and $R(b)$ is given by:
\begin{equation}
{
R(b,j)=-\int [dh]P(h)\int_{b}^{\infty}\,\dfrac{dt}{t}\, e^{ -Z(j,h) t}.
}
\end{equation} 
For large $b$, $R(b)$  decays exponentially, vanishing
for $b\rightarrow \infty$; 
therefore, the dominant contribution to the average free energy 
is given by the moments of the partition function of the model. 

To proceed, we assume the following probability distribution density $P(h)$ of 
the disorder field~$h$:
\begin{equation}
P(h) = p_{0}\,\exp\left\{-\frac{1}{2\,\varrho^{2}}\int\,d^{\mathrm{d}}x 
\int\,d^{\mathrm{d}}y\,\, h(x) F(x-y) h(y)\right\}{,}
\label{eq:probdis}
\end{equation}
where $\varrho$ is a positive parameter associated with the strength of the disorder, $p_{0}$ 
a normalization constant, and $F(x-y)$ defines the disorder correlation $
\mathbb{E}[{h(x)h(y)}] = \varrho^{2} \,  F(x-y)$. Then, integrating over the disorder, 
one can write the $k$-th moment of the partition function $\mathbb{E}\,[Z^{\,k}(j,h)]$  as:
\begin{equation} \label{eq:zeta}
\mathbb{E}\,[Z^{\,k}(j,h)] = \int\,\prod_{i=1}^{k}\left[d\phi_{i}^{k}\right]\,
e^{ - S_{\textrm{eff}}\left(\phi_{i}^{k},j_{i}^{k}\right)},
\end{equation}
where $S_{\textrm{eff}}\left(\phi_{i}^{k},j_{i}^{k}\right)$ is an effective action
for a multiplet of fields; that is, for the $k$-th moment, one has a 
multiplet of $k$ fields, $\phi^k = \left( \phi^k_1, \, \phi^k_2, \cdots, \, \phi^k_{k-1},\, \phi^k_k \right)^{T}$.
The form of $S_{\textrm{eff}}\left(\phi_{i}^{k},j_{i}^{k}\right)$ 
depends on the original model's action~$S(\phi)$. We define the model in the next section and discuss how it proves
an analog to topology fluctuations leading to the Euclidean wormholes discussed above. 

\section{Analog Model for Euclidean Wormholes: The Effective Action}
Our main idea is that the topology fluctuations
in the Euclidean path integral in Eq.~(\ref{grav-action})
can be effectively modeled by coupling a quenched disorder 
field to the matter field~$\phi$. In practice, one  
removes from the functional integral the functional measure 
of the metric and takes the disorder average of the corresponding 
free energy over ensembles of disorder realizations. 
Proceeding in this way, the Euclidean wormholes' 
effective action is readily identified. The topology fluctuation information 
is then effectively accounted for by the quenched disorder field.

We consider a scalar matter field~$\phi(x)$ coupled to a 
disorder field~$h(x)$ defined in~$\mathbb{R}^{\mathrm{d}}$. The partition 
function of this Euclidean field theory is given by
\begin{equation}  
    Z(h)= \int [d\phi] \, e^{ - S(\phi,h)},
\end{equation}
where the action $S(\phi,h)$ given by 
\begin{equation}
    S(\phi,h) = \int d^{\mathrm{d}}x \,
    \left[\frac{1}{2}\phi(x)(-\Delta + m^2)\phi(x) + h(x)\phi(x) \right].
     \label{action-orig}
\end{equation}
Since our interest is in global quantities, there is no need to couple an external source field~$j(x)$, as in Eq.~(\ref{eq:edisorderedgeneratingfunctional}). For this action, the effective actions $S_{\textrm{eff}}$ defining the $k$-th moment of the partition 
function in Eq.~(\ref{eq:zeta}) are given by:
\begin{equation}
S_{\textrm{eff}}(\phi_{i}^k) = \int d^{\mathrm{d}}x \int d^{\mathrm{d}}y
\sum_{i,j=1}^{k} \frac{1}{2}\phi_{i}^k(x) \, [G^{-1}(k)]_{ij}(x-y) \,  \phi_{j}^k(y) ,
\label{Seff1}
\end{equation}
where $[G^{-1}(k)]_{ij}(x-y)$ is the inverse of the two-point 
correlation function
\begin{equation}
[G^{-1}(k)]_{ij}(x-y) = \left[ \left(-\Delta + m^2\right)  
\delta^{(\mathrm{d})}(x-y)\delta_{ij} 
- \varrho^{2}\, F_{ij}(x-y) \right] ,
\label{G-1ij}
\end{equation}
where $F_{ij}(x-y)$ is the matrix with all entries equal to $F(x-y)$. The term 
proportional to~$\varrho^2$ comes from averaging over the random field~$h$ and contains a nonlocal contribution when $F(x-y)$ is not 
$\delta$-correlated in $(x,y)$. The nonlocal contribution is the analog to the 
nonlocal term in Eq.~(\ref{klebanov}). The first term in this last equation gives 
the bare contribution to the connected two-point correlation function even in the 
absence of disorder averaging, whereas the second term is normally a disconnected 
contribution but, due to the averaging, it became a connected 
contribution~\cite{livro4}. 

One can diagonalize the action $S_{\textrm{eff}}(\phi_{i}^k)$ in the manifold of the 
$k$-th multiplet of fields~$\phi^k_i$. The matrix $[G^{-1}(k)](x-y)$ is a $k \times k$ 
real and symmetric matrix. Then, by the spectral theorem of linear algebra, it can be
diagonalized by an orthogonal matrix. Denoting $[G^{-1}(k)]_D(x-y)$ the resulting diagonal matrix, it can be written as:
\begin{eqnarray}\label{eq:g-1}
    [G^{-1}(k)]_D = \left[
\begin{array}{cccc}
   G^{-1}_0(x-y) & 0& \cdots & 0 \\
   0 & G^{-1}_0(x-y) & \cdots & 0 \\
   \vdots & & \ddots & \vdots \\
   0 & \cdots &0& G^{-1}_0(x-y) - k \varrho^2 F(x-y) 
\end{array}\right]_{k\times k}
\end{eqnarray}
where $G^{-1}_0(x-y) =  ( - \Delta^2 + m^2)\delta^{({\mathrm{d}})}(x-y)$. Here, 
we made a choice concerning the position of the nonlocal term $ F(x-y)$; namely, 
we positioned it in the $k^{\mathrm{th}}$ element of the diagonal. Such a choice
is arbitrary; any other position is equally valid.

Now, the structure of the matrix in Eq.~(\ref{eq:g-1}) allows us to split the $k$-th moment 
effective action $S_{\textrm{eff}}(\phi_{i}^k)$ into two pieces, one with internal symmetry 
$\mathrm{O}(k-1)$, which is the original action replicated for $k-1$ fields and another 
that contains the nonlocal contribution. To show this, we denote the new fields obtained 
from the $\phi^k_i,\;i=1,\cdots,k$ from the diagonalizing transformation by 
\begin{equation}
\phi^k = \left(\begin{array}{c}
   \phi^k_1 \\[0.1true cm]
   \phi^k_2 \\[0.1true cm]
   \vdots \\[0.1true cm]
   \phi^k_{k-1}\\[0.1true cm]
   \phi^k_k
\end{array}\right) \longrightarrow 
\left(\begin{array}{c}
   \phi^k_1  \\[0.1true cm]
   \phi^k_2 \\[0.1true cm]
   \vdots \\[0.1true cm]
   \phi^k_{k-1} \\[0.1true cm]
   \phi
\end{array}\right).
\end{equation}
Since the transformation is orthogonal, the Jacobian in the path integral for  the
$k$-th moment of the partition function $\mathbb{E}\,[Z^{\,k}(h)]$ is unity: 
\begin{eqnarray}
    \prod_{i=1}^k \left[d\phi_i^{k}\right] &\longrightarrow& \prod_{a=1}^{k-1} 
    \left[d\phi_a^{k}\right][d\phi],
\end{eqnarray}
and one can write $\mathbb{E}\,[Z^{\,k}(h)]$ as :
\begin{equation}\label{eq:zk}
    \mathbb{E}\,[Z^{\,k}(h)] = \int \prod_{a=1}^{k-1} \left[d\phi_a^{k}\right]
    e^{-S_{O}(\phi^k_{a})} 
    \int [d\phi] \; e^{-S^{(k)}_\varrho(\phi)} .
\end{equation}
with 
\begin{align}
S_{O}(\phi^k_{a}) &= \int d^{\mathrm{d}}x \sum_{a=1}^{k-1}
\, \frac{1}{2}\phi_{a}^{k}(x) \, \left( -\Delta^2 + m^2\right)\, \phi_{a}^{k}(x),
\label{SOk-1}\\
S^{(k)}_{\varrho} (\phi) &= \int d^{\mathrm{d}}x\int d^{\mathrm{d}}y \; 
\frac{1}{2} \phi(x) \left[G^{-1}_0(x-y) - k \varrho^2 F(x-y) \right] \phi(x) .
\label{Srho}
\end{align} 

How does this result relate to the original works about Euclidean wormholes? First, as already mentioned, the Gaussian disorder correlation leads to a probability distribution similar to that obtained in Coleman's work Ref.~[\refcite{Coleman:1988tj}]. Second, instead of calculating the mean value of the partition 
function with the wormhole contributions integrated out with non-Gaussian 
distributions for the topological fluctuations, as done by Preskill~\cite{Preskill:1988na} 
and Gonz\'alez-D\'{\i}az~\cite{wormholesng}, here we are analyzing those effects 
on the disorder average of the free energy (the log of the partition function). As~mentioned after Eq.~(\ref{G-1ij}), such an 
average leads to connected correlation functions that would be 
disconnected correlation functions in the absence of disorder. This feature 
leads to the interpretation that the quenched field 
induces topology fluctuations, fluctuations that have ``propagators" 
associated with them, in Preskill's~\cite{Preskill:1988na} sense. 
Said differently, the disorder average of the free energy in 
Eq.~(\ref{eq:zk}) is actually a superposition of the contributions given 
by (infinitely) many universes connected by Euclidean wormholes. This is the 
analog to the proposal by Klebanov, Susskind and Banks~\cite{Klebanov:1988eh} 
that our universe was in a thermal bath with many (possibly infinite) universes. Finally, a similar interpretation of the average of the free energy 
was presented in a recent work by Okuyama~\cite{okuyama}, in which a different 
method was used to compute the average free energy. 

\begin{figure}[ht!]
     \centering
     \includegraphics[scale=0.35]{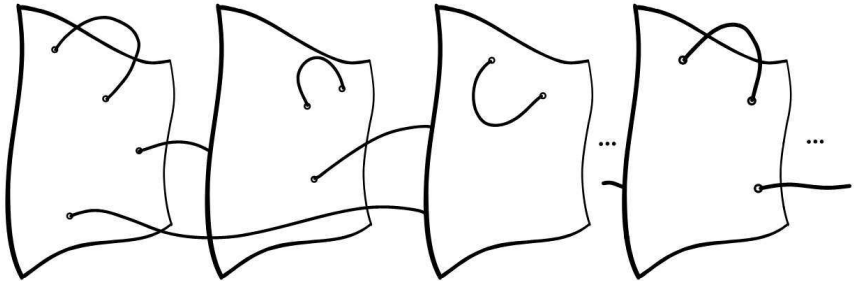}
     \caption{{\footnotesize Visualization of the topology fluctuations obtained from the disorder-averaged free energy of the model.}}
     \label{fig:branes}
 \end{figure}

It is important to point out that a single term in the series does not define a 
brane (universe); rather,
the brane interpretation applies only to the entire series. After the diagonalization and 
the redefinition of the fields in the functional space, a single term of the series has no direct interpretation 
at all. The entire series is needed to obtain physical quantities. Figure~\ref{fig:branes} provides a visualization
of our result, in that all topology fluctuations are, in fact, Euclidean wormholes. As evinced by Eq.~(\ref{Srho}),
we have two kinds of fluctuations: those that connect different branes (different universes), and those located on 
the same brane (same universe). 

A link with condensed matter physics is almost trivial. A disordered system at low temperatures, or for an anisotropic disorder leads to a model 
with the same mathematical structure regarding the nonlocality induced by
quantum gravity effects on matter fields. The series 
in Eq.~(\ref{eq:logz}) takes into account all possible configurations of the disorder. 
However, those configurations are not independent, since the disorder average is taken 
over the free energy, the generating functional of the connected correlation functions.

This concludes the formulation of the analog model. Physical quantities,
such as the dynamic and static structure factors, can be readily computed by 
using a mean-field approximation to obtain the necessary matter-field 
correlation functions. We recall that the static structure factor is proportional 
to the total intensity of light scattered by the fluid~\cite{2006v}. As such,
the effects of the disorder-induced nonlocality should leave signals on the 
scattered light.

 In summary,  we have proposed an analog model for Euclidean wormholes 
and topological fluctuation effects in a Riemannian space.
We aimed at modeling the effects of a quantum theory of gravitation on a matter
field. The idea of modeling the internal degrees of freedom by a random field has 
logical appeal and historical background. Although we based our derivations 
using a scalar field, the formalism can be easily adapted to other fields, 
such as vector and spinor fields.

\section*{Acknowledgments} 
The authors are gratful to L. Haiashi due the Fig. \ref{fig:branes} and fruitful discussions. This work was partially supported by Conselho Nacional de Desenvolvimento Cient\'{\i}fico e Tecnol\'{o}gico (CNPq), 
grants nos. 305894/2009-9 (G.K.) and 303436/2015-8 (N.F.S.), INCT F\'{\i}sica Nuclear e Apli\-ca\-\c{c}\~oes, grant no.
464898/2014-5  (G.K), and Funda\c{c}\~{a}o de Amparo \`{a} Pesquisa do Estado de S\~{a}o Paulo (FAPESP), grant no. 
2013/01907-0 (G.K). G.O.H thanks Coordena\c{c}\~ao de Aperfei\c{c}oamento de Pessoal de Nivel Superior (CAPES) for
a Ph.D. scholarship.  

\providecommand{\noopsort}[1]{}\providecommand{\singleletter}[1]{#1}%

\end{document}